\documentclass[11pt,a4paper]{article}
\usepackage{jcappub}

\usepackage{amsmath,amssymb,graphicx,multirow,dcolumn,bm,latexsym,soul,nicefrac,booktabs}
\usepackage{hyperref}
\hypersetup{
    colorlinks=true,
    linkcolor=blue,
    citecolor=blue,
    urlcolor=blue
}
\allowdisplaybreaks
\newcounter{RomanNumber}

\newcommand{\be}{\begin{equation}}
\newcommand{\ee}{\end{equation}}
\newcommand{\bea}{\begin{eqnarray}}
\newcommand{\eea}{\end{eqnarray}}

\usepackage{orcidlink} 
\usepackage{float}
\usepackage{xcolor}
\usepackage{subcaption}

\makeatletter
\gdef\@fpheader{} %
\makeatother

\begin{document}

\title{Constraining scalar-tensor theories from higher harmonics with GW230529}

\author[a]{Baoxiang Wang\,\orcidlink{0009-0003-3998-4609}}
\affiliation[a]{School of Physics and Technology, Wuhan University, Wuhan 430072, China}

\author[a,*]{Tao Yang\,\orcidlink{0000-0002-2161-0495}}
\emailAdd{yangtao@whu.edu.cn}

\abstract{
The Advanced LIGO and Virgo collaborations recently detected a gravitational wave event, GW230529\_181500, during the fourth observing run, which is most plausibly attributed to the merger of a neutron star and a black hole. This observation provides an opportunity to test a class of gravitational theories that deviate from general relativity. In such theories, additional terms contribute to the gravitational wave signal only in cases of asymmetric binaries. This study focuses on two scalar-tensor models within this class of theories: the Screened Modified Gravity and Brans-Dicke theory. These models have potential applications in areas such as dark matter, dark energy, cosmic inflation, and primordial nucleosynthesis. With the GW230529\_181500 and Bayesian Markov-chain Monte Carlo analyses, we derive a 90\% credible lower bound as $ \frac{\varphi_{\rm_{VEV}}}{M_{\rm Pl}}<1.7\times10^{-2}$ and $\omega_{\rm BD}>25.12$ by using dominant mode correction. Asymmetric binary systems usually have a significant mass ratio, in such cases, higher harmonic modes cannot be neglected. Our work considers higher harmonic corrections from scalar-tensor theories and provides a tighter constraint of $ \frac{\varphi_{\rm_{VEV}}}{M_{\rm Pl}}<1.5\times10^{-2}$ and $\omega_{\rm BD}>32.68$, with a 13.3\% and 30.1\% improvement respectively. Combining GW230529\_181500, GW200115 and GW190814 and including higher modes, the constraint is improved to $\frac{\varphi_{\rm_{VEV}}}{M_{\rm Pl}}<7.84\times10^{-3}$ and $\omega_{\rm BD}>123.75$. This is currently the strongest constraint from GWs, contingent upon GW190814 being an NSBH event.}


\maketitle

\section{Introduction}

The Advanced LIGO \cite{LIGOScientific:2014pky} and Virgo \cite{VIRGO:2014yos} collaborations recently concluded the first phase of the fourth observing run (O4a) at 16:00 UTC on January 16, 2024. Shortly after the start of O4a, aLIGO recorded a gravitational wave (GW) event, GW230529\_181500 (abbreviated as GW230529), which is most plausibly attributed to the merger of a neutron star and a black hole \cite{LIGOScientific:2024elc}.

In comparison to other methodologies such as laboratory and Solar System experiments, or binary pulsar and cosmological observations, GW events offer a powerful means of testing general relativity (GR) under strong-field or highly dynamical conditions \cite{LIGOScientific:2016lio,LIGOScientific:2018dkp,Yunes:2013dva,Will:2014kxa,Berti:2015itd,Berti:2018cxi}. Such scenarios are particularly useful for distinguishing GR from modified gravity theories. As the number of detected GW events grows, they become increasingly valuable tools for placing constraints on modified gravity frameworks, such as Einstein–scalar–Gauss–Bonnet (including the dilaton case, often referred to as Einstein–dilaton–Gauss–Bonnet) gravity  \cite{Sanger:2024axs, Yunes:2016jcc, Tahura:2019dgr, Nair:2019iur, Perkins:2021mhb, Wang:2023wgv, Gao:2024rel}, extra dimension \cite{Yunes:2016jcc}, time-varying gravitational constant G(t) \cite{Vijaykumar:2020nzc} and scalar-tensor theories  \cite{Yunes:2016jcc, Zhao:2019suc, Niu:2021nic, Damour:1992we,  sotiriou2014gravity, Bernard:2022noq, Sennett:2016klh, Trestini:2024zpi, Trestini:2024mfs, Bhattacharyya:2024aeq}.

Scalar-tensor theories \cite{Berti:2015itd}, one of the most natural extensions of GR, introduce additional scalar degrees of freedom that couple nonminimally to the metric tensor in the gravitational sector. Such scalar fields can arise through compactification mechanisms from higher-dimensional theories like string theory \cite{polchinski1998string}, Kaluza-Klein-like models \cite{duff1994kaluza}, or braneworld scenarios \cite{Randall:1999ee, Randall:1999vf}. Scalar-tensor frameworks offer a comprehensive platform to explore the phenomenology of fundamental theories and have been applied to study phenomena such as the Universe’s accelerated expansion \cite{Riazuelo:2001mg, Brax:2004qh, Kainulainen:2004vk}, inflation \cite{Clifton:2011jh, Burd:1991ns, Barrow:1990nv}, large-scale structure formation \cite{Brax:2005ew}, and primordial nucleosynthesis \cite{Coc:2006rt, Damour:1998ae, Larena:2005tu, Torres:1995je}. 
Here, we focus on two different models of scalar-tensor theories, i.e., the screened modified gravity (SMG) and the Brans-Dicke (BD) theory. Screened modified gravity theories, scalar degrees of freedom are introduced that can serve as dynamical dark energy components while evading stringent local gravity constraints through screening mechanisms \cite{Clifton:2011jh}. These mechanisms, such as the Chameleon \cite{Khoury:2010xi,Khoury:2003rn}, Vainshtein \cite{Vainshtein:1972sx,Babichev:2013usa}, and Symmetron \cite{Hinterbichler:2010es} mechanisms, enable the suppression of deviations from GR in high-density or small-scale environments, thereby ensuring consistency with precise laboratory and Solar System tests. On cosmological scales, however, the scalar field can remain unscreened and dynamically drive the late-time acceleration of the Universe, offering a unified framework for modified gravity and dark energy phenomenology. Extensive studies have explored the theoretical consistency, astrophysical consequences, and cosmological signatures of Screened modified gravity theories \cite{Joyce:2014kja,Clifton:2011jh,Brax:2012bsa,Fischer:2024eic}. For Brans-Dicke  theory \cite{Brans:1961sx}, originally developed by Jordan, Fierz, Brans, and Dicke, which incorporates Mach’s principle by promoting the gravitational constant to a dynamical scalar field. The interaction strength is governed by a dimensionless coupling parameter $\omega_{\rm BD}$, with general relativity recovered in the limit $\omega_{\rm BD} \to \infty$. Extensive observational efforts have placed bounds on $\omega_{\rm BD}$ using various systems. For instance, binary pulsar timing has constrained $\omega_{\rm BD} > 25000$ from the orbital decay of PSR J1738+0333 \cite{Freire:2012mg}, while the most stringent limit to date, $\omega_{\rm BD} > 40000$, arises from the Shapiro time delay measured by the Cassini mission in the Solar System \cite{Bertotti:2003rm, Will:2014kxa}.

Meanwhile, there are also many other applications for GW in constraints. In scalar-tensor  theory, the waveform are influenced by neutron star–black hole (NSBH) binaries because the deviations from GR in the waveform are related to the ‘sensitivities’ of the binary components \cite{Freire:2012mg}. Therefore, current limits on scalar-tensor theory are rely on NSBH systems, such as using space-based GW detectors in future simulations \cite{Berti:2004bd, Yagi:2009zm, Scharre:2001hn, Jiang:2021htl, Gao:2022hsn}. In analyses based on observed gravitational wave events, Rui Niu et al. analyzed the GW200115 event and derived the bounds $\frac{\varphi_{\mathrm{VEV}}}{M_{\mathrm{Pl}}} < 1.8 \times 10^{-2}$ and $\omega_{\mathrm{BD}} > 40$  \cite{Niu:2021nic} using Bayesian inference implemented with the open-source package Bilby \cite{Ashton:2018jfp}. Tan et al. conducted a joint analysis of GW200115 and GW190814 and found that $\frac{\varphi_{\mathrm{VEV}}}{M_{\mathrm{Pl}}} < 8.32\times 10^{-3}$ and $\omega_{\mathrm{BD}} > 110.55$  \cite{Tan:2023fyl}. Similarly, using the breathing scalar polarization, Takeda et al. reported a constraint of $\omega_{\rm BD}>81$ \cite{Takeda:2023wqn}. 

The observational sources currently available for constraining scalar-tensor theory are largely limited to asymmetric binary systems, including NSBH, white dwarf–neutron star, and white dwarf–black hole binaries. GW230529 provides a valuable new data point for this purpose. Our work incorporates higher harmonic modes into the waveform, as such effects cannot be neglected in asymmetric binary systems with relatively large mass ratios. Specifically, we employ the open-source software PyCBC \cite{Biwer:2018osg} and Bayesian inference methods \cite{Thrane:2018qnx, Smith:2021bqc, Lyu:2022gdr} within the parameterized post-Einsteinian (ppE) framework \cite{Yunes:2009ke, Yunes:2010qb, Mirshekari:2011yq, Tahura:2018zuq} to analyze GW data. Using the IMRPhenomXHM waveform model, our analysis of GW230529 data yields a constraint on the scalar-tensor coupling parameter of $\frac{\varphi_{\rm_{VEV}}}{M_{\rm Pl}}<1.7\times10^{-2}$ and $\omega_{\rm BD}>25$. By including higher-mode effects we obtain constraints of $\frac{\varphi_{\rm_{VEV}}}{M_{\rm Pl}}<1.5\times10^{-2}$ and $\omega_{\rm BD}>32.68$. The former represents a 13.3\% and 30.1\% improvement over the dominant mode case. Meanwhile, using the observations of GW230529, GW200115, and GW190814, which is assumed to be an NSBH event, the constraints improve further, such that $\frac{\varphi_{\rm_{VEV}}}{M_{\rm Pl}}<7.84\times10^{-3}$ and $\omega_{\rm BD}>123.75$. This is currently the strongest constraint utilizing GWs.

The structure of this paper is as follows: Section \ref{sec:model}, we explain the construction of the waveform models used in this paper. Section \ref{sec:Bayes} describes the Bayesian inference method used for data analysis. Section \ref{sec:result} presents the main results, and Section \ref{sec:sumout} concludes the paper with a summary and outlook. Throughout, we adopt the convention $G = c = 1$.

\section{Waveform}\label{sec:model}
In this section, we introduce the ppE formalism and show how the waveforms in screened modified gravity and Brans–Dicke theories map into it.

\subsection{Parameter post-Einsteinian framework}

The coalescence of compact binary systems can be categorized into three sequential phases: inspiral, merger, and ringdown. In the inspiral phase, the two compact objects remain well separated, moving at relatively low velocities, which permits the application of the post-Newtonian (PN) approximation to accurately model waveforms for systems with small mass ratios \cite{Blanchet:2013haa}.
In our analysis, we employ the parameterized ppE framework—a general parametric method introduced by Yunes and Pretorius to characterize systematic deviations of PN waveforms in various modified gravity (MG) theories from those predicted by GR \cite{Yunes:2009bv}. When considering only the leading-order correction, the inspiral waveform within the ppE formalism consists of the standard GR contribution supplemented by additional terms encoding possible modifications \cite{Yunes:2009ke}:

\begin{equation}\label{ppe_h_com}
\tilde{h}(f)=\tilde{h}_{\rm GR}(f) \,(1+\alpha\, u^a)e^{i \beta u^b}\, ,
\end{equation}
where $f$ is the frequency, and $(\alpha u^a, \beta u^b)$ are the amplitude and phase corrections from modified gravity. $b=2{\rm~ PN}-5$ and $a=b+5$ are the ppE order parameters, and the $\rm~ PN$ in this equation symbolizes the PN order, $u={(\pi \mathcal{M} f)}^{1/3}$ is the reduced frequency of the inspiral that is proportional to the relative velocity of the binary,
$\mathcal{M}=\eta^{3/5}M$ is the chirp mass,  $M=m_1+m_2$ is the total mass, $\eta=m_1m_2/(m_1+m_2)^2$ is the symmetrical mass ratio, $m_1$ and $m_2$ are the major and minor masses, respectively.

One can map Eq.\eqref{ppe_h_com} to the basis of spin-weighted spherical harmonics  $ Y_{-s}^{\ell m}(\iota, \varphi) $ to obtain the harmonic decomposition of ppE waveform \cite{Mezzasoma:2022pjb}:
\begin{align}
\tilde{h}_{\rm ppE}(f) &= \sum_{\ell m} \tilde{h}_{\ell m}^{\rm ppE}(f) \,, \nonumber \\
\tilde{h}_{\ell m}^{\rm ppE}(f) &= \tilde{h}^{\rm GR}_{\ell m}(f) \, (1+\alpha_{\ell m}\, u^a_{\ell m}) e^{i \beta_{\ell m} u^{b_{\ell m}}} \,. \label{eq:phase3}
\end{align}

In this paper, the GR waveform will be produced using IMRPhenomXHM, and the higher harmonics modes $(\ell,|m|)=\{(2,2),(2,1),(3,3),(3,2),(4,4) \}$ are considered. Both $ \{ a_{\ell m},\alpha_{\ell m},b_{\ell m} ,\beta_{\ell m} \}$ can be different for different values of $\ell$ and $m\,$.

We focus on phase corrections, since Tahura et al.~\cite{Tahura:2019dgr} validated the common approach of considering only phase corrections. 
They showed that including amplitude corrections together with phase corrections changes the inferred bounds by at most $\sim 4\%$ across several modified gravity theories. 
Thus, following the standard practice in the literature, we adopt the phase-only ppE waveform,
which reduces to

\begin{equation}
	\tilde{h}_{\ell m}^{\rm ppE}(f)= \tilde{h}_{\ell m}^{\rm {GR}}(f) e^{i\beta_{\ell m} u^{b_{\ell m}}}.
\label{wave_ppe_pha}
\end{equation} 
The phase correction with higher harmonics in \eqref{wave_ppe_pha} can be derived as \cite{Mezzasoma:2022pjb}:
\begin{equation}\label{ppe_b}
	\beta_{\ell m} = \beta_m , \qquad b_{\ell m} = b ,
\end{equation}
with 
\begin{equation}\label{ppe_beta}
	\beta_{m} = \left(\frac{2}{m}\right)^{b/3-1}\beta_{22} .
\end{equation}
It is worth noting that this differs from the waveform treatment adopted by S\"anger et al., who performed a theory agnostic analysis by introducing generic deformations, and by construction their prescription corresponds within the ppE framework to $\beta_{\ell m} = (m/2)\,\beta_{22}$ \cite{Sanger:2024axs}. In addition, this also differs from the waveform adopted by Gao et al., which treats $\beta_{\ell m}$ uniformly across modes \cite{Gao:2024rel}.

\subsection{Waveform in scalar-tensor theory}
In this work we consider a class of scalar-tensor theories, which can be described by the \cite{Chiba:1997ms}
\begin{equation}
\begin{aligned}
S= & \frac{1}{16 \pi} \int d^4 x \sqrt{-\tilde{g}}\left[\tilde{R}-2 \tilde{g}^{\mu \nu}\left(\partial_\mu \varphi\right)\left(\partial_\nu \varphi\right)-V(\varphi)\right]  \\
&+S_M\left[\Psi, \phi(\varphi) \tilde{g}_{\mu \nu}\right] \,.
\label{ST_action}
\end{aligned}
\end{equation}
Here, $\tilde{g}^{\mu\nu}$ and $\tilde{g}$ denote the Einstein-frame metric and its determinant, respectively. $\tilde{R}\equiv \tilde{g}^{\mu\nu} \tilde{R}^{\mu\nu}$ is the Ricci scalar. 
$\varphi$ denotes the scalar field, and $V(\varphi)$ is its potential, which gives the field an effective mass. For Brans–Dicke theory, the potential term $V(\varphi)$ is not considered. $\Psi$ denotes the matter fields and $\phi(\varphi)$ is the conformal coupling function. 
In scalar-tensor theories, compact binary systems exhibit a misalignment between the centers of gravitational binding energy and inertial mass. This asymmetry leads to a time-varying dipolar structure, which radiates additional energy through scalar emission \cite{Will:1994fb}. In our analysis, we adopt a waveform model incorporating the leading order correction to general relativity, characterized by a dipole contribution to the gravitational wave phase for the dominant mode \cite{Will:1994fb,Tahura:2018zuq,Zhang:2017sym,Liu:2020moh}, 
\be \label{dipole_term_1}
h(f) = h_{\rm GR}(f)\exp\left[ i \frac{3}{128\eta} \varphi_{-2} (\pi  M f)^{-7/3} \right].
\ee
Here $\eta=m_1m_2/M^2$ is the symmetric mass ratio. $\varphi_{-2}=-5(\alpha_{\rm A}-\alpha_{\rm B})^2/168$, where $\alpha_{\rm A} \text{ and } \alpha_{\rm B}$ are scalar charges of the components of the binary. 
For black holes, the no-hair theorem prevents them to acquire scalar charges. In many scalar-tensor theories including the models considered in this work, where the no-hair theorem can be applied, scalar charges of black holes are 0 \cite{Hawking:1972qk,Bekenstein:1995un,Sotiriou:2011dz}.

For screened modified gravity theory, we only consider the leading order modification and replace $\alpha_{\rm A}-\alpha_{\rm B}$  as
$\varphi_{\rm_{VEV}}/({M_{\rm Pl}\Phi_A})$. Where $M_{\rm Pl}=\sqrt{1/8\pi G}$ is the reduced Planck mass, $\varphi_{\rm VEV}$ is the vacuum expectation value of the scalar field, and $\Phi_A=Gm/R$ is the surface gravitational potential of the object $A$.
In conjunction with Eq.\eqref{ppe_h_com}, the dominant mode correction of these ppE parameters in screened modified gravity theory  can be written as \cite{Freire:2012mg,Tahura:2018zuq, Liu:2020moh}:
\begin{eqnarray}
\beta_{22}^{\rm SMG} &=&\frac{-5}{7168}\eta ^{2/5} \varphi_{\rm VEV}^2/({M_{\rm Pl}\Phi_A})^2 .
\end{eqnarray}

For Brans-Dicke theory, where one can replace $(\alpha_{\rm A}-\alpha_{\rm B})^2$  as $2  (s_{\rm A}-s_{\rm B})^2/(2+\omega_{\rm BD})$ \cite{Freire:2012mg}. Here, $s_{\rm A}$ denotes the sensitivity of the $A$th body, which approximately equals its compactness (about 0.5 for black holes and $\sim 0.2$ for neutron stars).
Combining the insights from Eq.\eqref{ppe_h_com}, the dominant mode correction of these ppE parameters in BD theory  can be written as \cite{Freire:2012mg,Tahura:2018zuq, Liu:2020moh}:
\begin{eqnarray}
\beta_{22}^{\rm BD} &=&\frac{-5}{3584}\eta ^{2/5} (s_{\rm A}-s_{\rm B})^2/(2+\omega_{\rm BD}).
\end{eqnarray}
The leading order modification in the GW waveform introduced by scalar-tensor theory starts at the $-1$PN order, corresponding to $b^{\rm SMG}=b^{\rm BD}=-7$. And by combining Eqs.\eqref{ppe_b} and Eq.\eqref{ppe_beta}, we can derive $\beta_{\ell m}^{\rm SMG}$ and $\beta_{\ell m}^{\rm BD}$.

\section{Bayesian inference}\label{sec:Bayes}

The ppE waveform formalism introduces a parametric model to describe possible deviations from general relativity in gravitational wave signals. This model can be incorporated into Bayesian inference frameworks to extract physical parameters or place constraints on modified gravity theories from observed waveforms \cite{Thrane:2018qnx, Smith:2021bqc, Lyu:2022gdr}. Bayesian inference is based on Bayes’ theorem, which relates the conditional and marginal probabilities of random events. Given a model hypothesis $\mathcal{H}$ and a set of observational data $d$, the posterior probability distribution of the model parameters $\boldsymbol{\vartheta}$ can be obtained via Bayes’ theorem:
\begin{equation}
\label{bayes_infre}
p(\boldsymbol{\vartheta} | d, \mathcal{H})
=  \frac{p(\boldsymbol{\vartheta} | \mathcal{H})  p(d | \boldsymbol{\vartheta}, \mathcal{H}) }{p(d | \mathcal{H})},
\end{equation}
where $p(\boldsymbol{\vartheta} | \mathcal{H})$ represents the prior distribution over parameters under the hypothesis $\mathcal{H}$, and $p(d | \boldsymbol{\vartheta}, \mathcal{H})$ is the likelihood function, which quantifies the probability of observing the data $d$ assuming the parameters $\boldsymbol{\vartheta}$.
The likelihood is not derived directly from the waveform model itself, but rather constructed by statistically comparing the observed data with theoretical waveform predictions, such as those generated under the parameterized post-Einsteinian (ppE) formalism while accounting for the noise properties of the detectors. Typically, the noise is modeled as stationary and Gaussian. The denominator $p(d | \mathcal{H})$ is known as the Bayesian evidence, which serves as a normalization factor and encapsulates the overall plausibility of the hypothesis $\mathcal{H}$ given the data. 

\begin{table}[htbp]
\centering
\small %
\begin{tabular}{c|l}
\hline\hline
$m_1$             & Mass of the major component   \\
$m_2$             & Mass of the minor component \\
$\chi_{1z}$             & Aligned or anti-aligned spins of the major component \\
$\chi_{2z}$             & Aligned or anti-aligned spins of the minor component \\
$\alpha_S$        & Right ascension of source location \\
$\delta$          & Declination of the source location \\
$\psi$            & Polarization angle \\
$\iota$           & Inclination angle \\
$\phi_{\rm ref}$  & Phase at the reference frequency  \\
$t_c$              & Coalescence time \\
$D_L$              & Luminosity distance \\
$\frac{\varphi_{\rm_{VEV}}}{M_{\rm Pl}},  \omega_{\rm_{BD}}$ & Parameters to be bounded of SMG and BD separately. \\
\hline\hline
\end{tabular}
\caption{\footnotesize
Summary of physical parameters and their meaning.}
\label{parameters_meaning}
\end{table}

Under the assumption of stationary Gaussian noise, the logarithm of the likelihood function takes the following form:
\begin{equation}
\log p(d|\boldsymbol{\vartheta}, \mathcal{H}) = \log \bar{\alpha} - \frac{1}{2} \sum_k \left< d_k - h_k(\boldsymbol{\vartheta}) \middle| d_k - h_k(\boldsymbol{\vartheta}) \right>,
\end{equation}
where $\log \bar{\alpha}$ is a normalization constant, the index $k$ labels individual detectors in the interferometer network, and $d_k$ and $h_k(\boldsymbol{\vartheta})$ denote the observed data and template waveform, respectively, for the $k$-th detector. The inner product is defined in the frequency domain as
\begin{equation}
\label{bayes_inner}
\left<a(t)\middle|b(t)\right> = 2 \int_{f_{\rm low}}^{f_{\rm high}} \frac{\tilde{a}^*(f) \tilde{b}(f) + \tilde{a}(f) \tilde{b}^*(f)}{S_n(f)} df,
\end{equation}
where $S_n(f)$ denotes the one-sided power spectral density (PSD) of the detector noise, and $(\tilde{a}(f), \tilde{b}(f), \tilde{a}^*(f), \tilde{b}^*(f))$ are the Fourier transforms of the time-domain signals $a(t)$ and $b(t)$, respectively. $f_{\rm low}$ denotes the low-frequency cutoff of the gravitational wave data, $M f_{\mathrm{max}} = 0.018$ represents the intermediate transition frequency of the inspiral stage, as specified by the phenomenological model \cite{Sampson:2013jpa, LIGOScientific:2019fpa, Khan:2015jqa}.

In our parameter estimation analysis, we sample over a total of 12 intrinsic and extrinsic parameters, including the parameter $\frac{\varphi_{\rm_{VEV}}}{M_{\rm Pl}}$ in the screened modified gravity theory and Brans–Dicke coupling constant $\omega_{\rm BD}$. The physical interpretations of each parameter are summarized in Table~\ref{parameters_meaning}.

\section{Results}\label{sec:result}
In scalar–tensor theories, corrections to the GW phase relative to GR depend on the difference in sensitivities of the compact binary components. Consequently, for symmetric binaries with equal sensitivities, the scalar–tensor correction vanishes, and the waveform is indistinguishable from GR predictions~\cite{Freire:2012mg}.

Recently, the LIGO-Virgo-KAGRA (LVK) collaboration reported the detection of GW230529, a GW signal most plausibly originating from the coalescence of a NSBH. Leveraging this event, we constrain scalar–tensor theories by performing Bayesian inference over the parameter space $\boldsymbol{\vartheta}$.

\begin{figure}[htbp]
    \centering
    \includegraphics[width=0.8\textwidth]{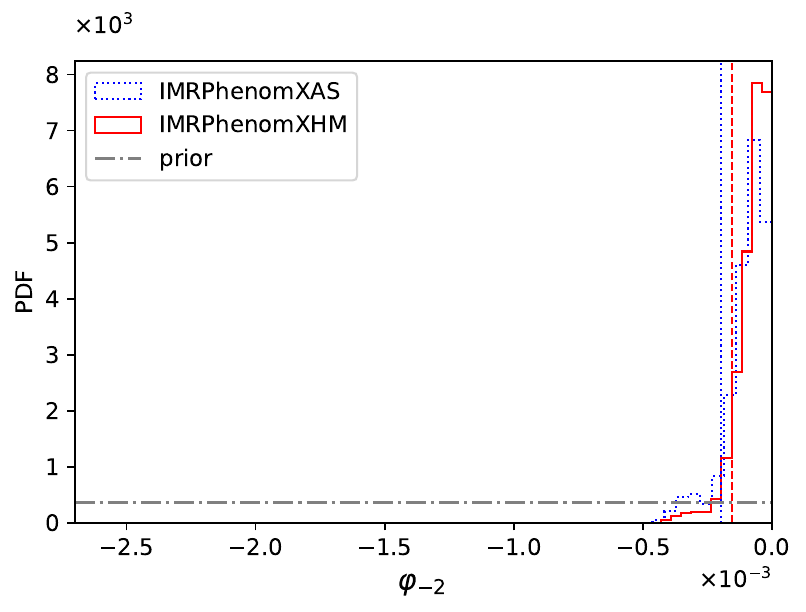}
    \caption{Probability density of Bayesian posterior distributions on $\varphi_{-2}$. The blue dashed line corresponds to the 90\% credible constraint using the dominant mode correction, while the red solid line denotes the 90\% credible constraint using the correction including higher modes. The gray dash-dotted line represents the prior on $\varphi_{-2}$.
    }
    \label{fig_phi_-2}
\end{figure}

The analysis is carried out using the open-source PyCBC software package~\cite{Biwer:2018osg}, employing MCMC sampling with the emcee\_pt sampler~\cite{Foreman-Mackey:2012any}, initialized with 200 walkers. To mitigate contamination from low-frequency noise, we adopt a lower frequency cutoff of $f_{\rm low} = 20$~Hz. A data segment of 64 seconds centered around GW230529 is used for parameter estimation. As for prior choices, we assume a uniform distribution for $1/(\omega_{\rm BD} + 2)$ in the range $[0, 0.5]$, and for $\varphi_{-2}$ in the interval $[-0.0027, 0]$ (Some theory-agnostic analyses adopt priors spanning both positive and negative values; for a comparison with negative-only priors and the resulting impact on the posterior, see Appendix~\ref{robustness}). Spin components $\chi_{1z}$ and $\chi_{2z}$ are aligned or anti-aligned with the orbital angular momentum, and are independently drawn from a uniform distribution in the range $[-0.99, 0.99]$.

\begin{figure}[htbp]
    \centering
    \includegraphics[width=0.75\textwidth]{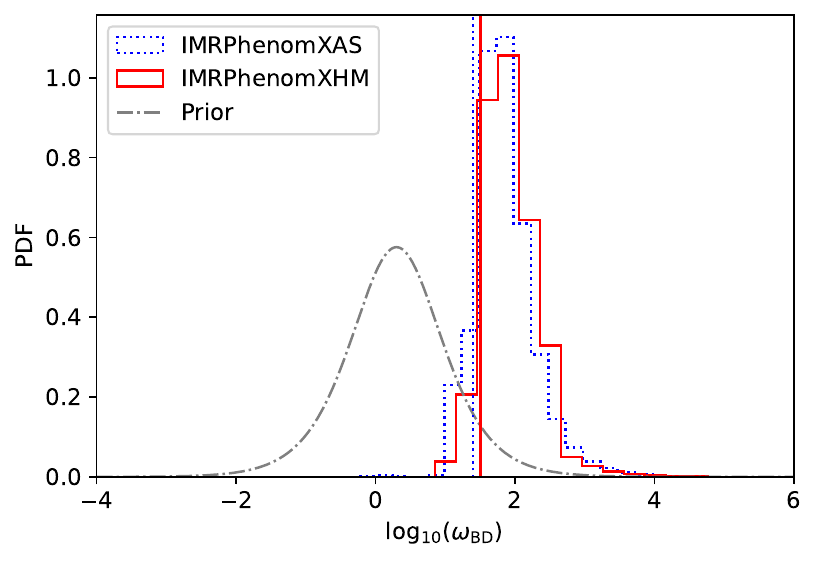}
    \caption{Probability density of Bayesian posterior distributions on $\omega_{\rm_{BD}}$. The blue dashed line corresponds to the constraint on BD at 90\% credible level by using the dominant mode correction, while the red solid line represents the constraint on BD with 90\% probability by using the correction including higher mode. The gray dash-dotted line represents the prior on $\omega_{\rm_{BD}}$. We use a logarithmic scale due to the posterior distribution of $\omega_{\rm BD}$ spanning several orders of magnitude.}
    \label{fig_bd}
\end{figure}

Our analysis focuses on the inspiral regime and neglects tidal interactions in NSBH systems~\cite{LIGOScientific:2021qlt}, and we employ the IMRPhenomXHM waveform model from the LALSimulation package~\cite{collaboration2018ligo} for both the screened modified gravity and Brans–Dicke hypotheses. 
These phenomenological models operate in the frequency domain and incorporate the effects of non-precession and higher-order multipole moments.

\begin{figure}[htbp]
    \centering
    \includegraphics[width=0.75\textwidth]{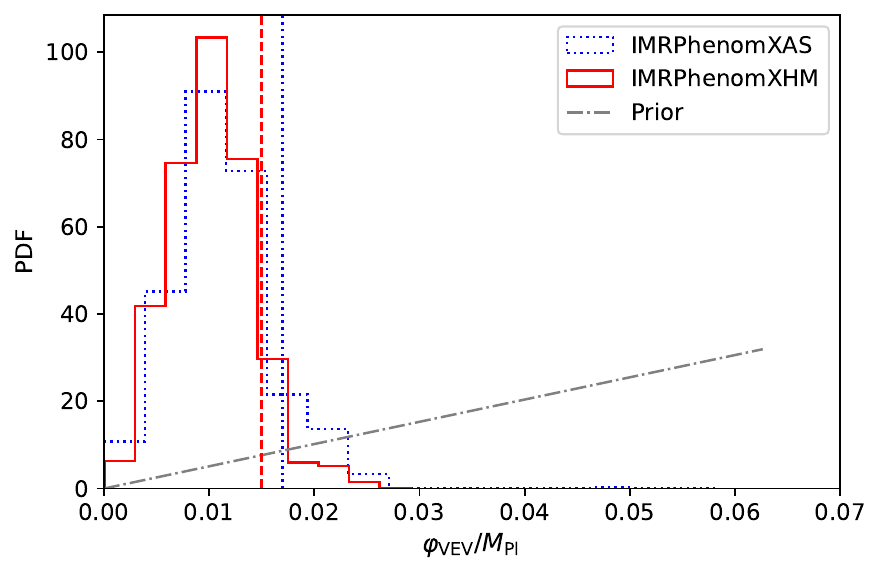}
    \caption{Probability density of Bayesian posterior distributions on ${\varphi_{\rm_{VEV}}}/{M_{\rm Pl}}$. The blue dashed line corresponds to the constraint at a 90\% credible level by using the dominant mode correction, while the red solid line stands for the constraint at the 90\% confidence level by using the correction including the higher mode. The gray dash-dotted line represents the prior on ${\varphi_{\rm_{VEV}}}/{M_{\rm Pl}}$.}
    \label{fig_varphi_VEV}
\end{figure}

We present the resulting constraints on $\varphi_{-2}$ in scalar–tensor gravity in Fig.~\ref{fig_phi_-2}. For GW230529, we obtain a 90\% credible lower bound of $\varphi_{-2} > -1.98 \times 10^{-4}$ using the dominant mode (DM) correction only.
It is consistent, in order of magnitude, with S\"anger et al.'s ESGB constraint based on the dominant mode only~\cite{Sanger:2024axs}.
When higher-order modes HM are included, the constraint improves to $\varphi_{-2} > -1.54 \times 10^{-4}$, corresponding to a 22.22\% enhancement. Corresponding constraints on the Brans–Dicke coupling parameter $\omega_{\rm BD}$ are shown in Fig.~\ref{fig_bd}. For GW230529, the dominant mode analysis yields a 90\% lower bound of $\omega_{\rm BD} > 25$. Including higher-order modes improves the bound to $\omega_{\rm BD} > 32.68$, a 30.1\% enhancement. For sreened modified gravity theory, we adopt neutron star parameters $m = 1.4 M_\odot$ and $R = 10$km. Using the same Bayesian pipeline and the GW230529 signal, we constrain the scalar field vacuum expectation value $\varphi_{\rm VEV}$. We find that $\varphi_{\rm VEV}/M_{\rm Pl} < 1.7 \times 10^{-2}$ using the dominant mode, and $\varphi_{\rm VEV}/M_{\rm Pl} < 1.5 \times 10^{-2}$ when higher-order modes are included.

It is worth emphasizing that the constraint on $\varphi_{\rm VEV}/M_{\rm Pl}$ can be interpreted within a broad class of screened modified-gravity frameworks, including Vainshtein-type and K-essence–type screening (also referred to as kinetic screening or k-mouflage). In these scenarios, $\varphi_{\rm VEV}$ represents the asymptotic vacuum value of the scalar field, which sets the strength of the fifth force before screening becomes active. Our bound therefore constrains the effective scalar charge difference between the binary constituents at orbital separations of order $10$–$100\,\mathrm{km}$ probed in the late inspiral.

For Vainshtein-type screening~\cite{Babichev:2013usa,deRham:2012fw}, the suppression of scalar charges is controlled by the Vainshtein radius, 
$r_V \sim \left(\frac{GM}{\Lambda^3}\right)^{1/3}$, 
as in the cubic Galileon case, where $\Lambda$ denotes the strong-coupling scale. For typical parameter values considered in Galileon and massive gravity models, compact objects are well within their Vainshtein radii, so dipole radiation is highly suppressed. Our constraint therefore limits the $\varphi_{\rm VEV}$, but it does not yield a unique bound on $r_V$ without committing to a specific model. For K-essence screening (kinetic screening/k-mouflage)~\cite{Brax:2012jr,Babichev:2009ee}, the effective scalar charge depends on the functional form of the kinetic term $K(X)$ and its derivatives. While our constraint excludes large bare couplings, a quantitative mapping to a ``screening radius’’ or scalar sound speed again requires specifying the detailed form of $K(X)$. In this sense, our result is generic in ruling out sizable effective scalar charges at the orbital scales probed by GW230529, across different screening mechanisms. However, translating this bound into a constraint on a screening radius or other model-specific parameter is only possible once a concrete Lagrangian is chosen.

We also consider a combined analysis of the three NSBH events, GW230529, GW200115 and GW190814. 
For the stacking procedure \cite{Isi:2019asy, Isi:2022cii,Perkins:2021mhb}, we follow the methodology described in Perkins et al.~\cite{Perkins:2021mhb}.
Specifically, we implemented both of the approaches outlined there. 
In the first method, for each event we construct a histogram of the MCMC posterior samples and then obtain a smooth analytic representation of the posterior by applying kernel density estimation (KDE).  We then multiply the resulting KDEs pointwise across events to form the joint posterior.
In the second, more direct method, we align the binning across events, construct histograms of the posterior samples, and then multiply these histograms bin by bin, which is equivalent to multiplying the marginalized likelihoods of independent events. 
In both cases, mild smoothing is applied when evaluating confidence intervals in order to mitigate numerical fluctuations at the distribution tails. 
Throughout this work we quote the more conservative of the two results.

Due to convergence issues in the dominant mode only case for GW190814 \cite{Tan:2023fyl}, We cannot report the combined result for this using the dominant mode. Therefore, we report only the result using higher-order modes for the combined dataset: $\varphi_{-2} > -4.26 \times 10^{-5}$, demonstrating substantial improvement. We follow the same principle, and quote the more conservative constraint of the two methods.
In the combined analysis of the three NSBH events incorporating higher modes, the constraint strengthens significantly to $\omega_{\rm BD} > 123.75$. 
Similarly, Considering the combination, we obtain a tighter constraint as well $\varphi_{\rm VEV}/M_{\rm Pl} < 7.84 \times 10^{-3}$. All the above resulting bounds are summarized in Table~\ref{result_summary}.
\begin{table*}[h]
\centering
\small %
\begin{tabular}{lccc}
\hline\hline
 & \multicolumn{2}{c}{GW230529}    & Combination (GW230529,GW200115,GW190814) \\
\hline
 & Dominant mode (DM) & Higher mode (HM) & Higher mode (HM) \\
\hline
$\varphi_{-2}$ & $-1.98\times 10^{-4}$ & $-1.54\times 10^{-4}$ & $-4.26\times 10^{-5}$ \\
$\frac{\varphi_{\rm VEV}}{M_{\rm Pl}}$ & $1.7\times 10^{-2}$ & $1.5\times 10^{-2}$ & $7.84\times 10^{-3}$ \\
$\omega_{\rm BD}$ & 25.12 & 32.68 & 123.75 \\
\hline\hline
\end{tabular}
\caption{Summary of constraints on scalar–tensor gravity at a 90\% credible level.}
\label{result_summary}
\end{table*}

\section{Conclusion and discussion}\label{sec:sumout}
In this work, we revisited gravitational wave generation in scalar–tensor theories and outlined the construction of waveform models within the parameterized post-Einsteinian (ppE) framework. Using Bayesian inference with MCMC sampling implemented via the PyCBC software package, we analyzed the event GW230529 and derived constraints on screened modified gravity and Brans–Dicke theory. Motivated by the importance of higher-order multipole modes in asymmetric binaries with large mass ratios, we employed the IMRPhenomXHM waveform model to incorporate such effects. Our analysis yields bounds of $\varphi_{\rm VEV}/M_{\rm Pl} < 1.7 \times 10^{-2}$ and $\omega_{\rm BD} > 25$ using only the dominant mode, which improve to $\varphi_{\rm VEV}/M_{\rm Pl} < 1.5 \times 10^{-2}$ and $\omega_{\rm BD} > 32.68$ when higher modes are included, representing 13.3\% and 30.1\% enhancements, respectively. A joint analysis of GW230529, GW200115, and GW190814 (assumed to be a neutron star–black hole event) further tightens the constraints to $\varphi_{\rm VEV}/M_{\rm Pl} < 7.84 \times 10^{-3}$ and $\omega_{\rm BD} > 123.75$, currently the most stringent limits obtained from gravitational wave observations.

Recent theory-agnostic analyses have investigated potential GR deviations in the GW230529 event by employing higher-mode waveforms with generic phase corrections. However, the waveform modifications adopted in our scalar–tensor gravity analysis differ fundamentally from those used in such agnostic approaches. In particular, our higher-mode waveform model incorporates theory-specific phase corrections derived from the structure of scalar–tensor theories, as given in Eqs.(2.4)–(2.5). This key difference in waveform construction implies that the results from existing agnostic analyses cannot be directly mapped onto scalar–tensor gravity in the context of our study.

It should be noted that, unlike in some other studies where the inspiral cutoff frequency is treated as a fixed value, in our analysis it varies across different MCMC realizations according to the sampled total mass, following the phenomenological prescription $M f_{\rm max}=0.018$.
This treatment yields a more consistent and realistic implementation in Bayesian parameter estimation, and the resulting constraints, though slightly more conservative, are also more faithful to the actual data.
To assess the impact of this choice, we also tested an alternative prescription in which the cutoff frequency is fixed. 
In this case, the constraint obtained for GW230529 is $\varphi_{-2} > -8.5 \times 10^{-5}$ at the 90\% credible level, which differs by about a factor of two compared to the result obtained with the varying-cutoff prescription.
By stacking multiple events with this result, one can obtain $\varphi_{-2} > -3.4 \times 10^{-5}$, corresponding to $\omega_{\mathrm{BD}} > 156$ and $\varphi_{\mathrm{VEV}}/M_{\mathrm{Pl}} < 7 \times 10^{-3}$.

We reconstructed the PSD directly from the strain data using the Welch method, rather than adopting the event-specific PSD from LVK posterior releases. For robustness, we also performed a comparison between the two PSD prescriptions and found that the resulting constraints differ by less than 5\%, demonstrating consistency between the two approaches (see Appendix~\ref{robustness} for details).

In the LVK analysis of GW230529 \cite{LIGOScientific:2024elc}, waveform models incorporating both higher modes and spin precession were employed, alongside models including explicit tidal corrections. Across these analyses, there was no significant evidence for measurable precession or tidal effects. Importantly, including these additional degrees of freedom did not significantly alter the constraints on non-GR parameters. This justifies our use of IMRPhenomXHM as a clean comparison waveform, which allows us to focus on isolating the role of higher modes in constraining scalar–tensor gravity with GW230529 without introducing systematic uncertainties from precession or tidal effects.

More broadly, the different sources of systematic error in GW inference can be broadly classified into biases induced by inaccuracies in the data, noise, priors, waveforms, astrophysics, nuclear physics, and the underlying theory of gravity. The waveform systematics specifically include spin precession, high-PN corrections, eccentricity, tidal effects, and higher modes \cite{Chandramouli:2024vhw, Owen:2023mid, Read:2023hkv, Wade:2014vqa, Favata:2013rwa}. These factors are likely to affect tests of GR. Therefore, for scalar–tensor theories, as future implementations of new waveform models incorporating additional effects in higher harmonics become available, it will be important to further investigate these factors for sources other than GW230529-like events, where no evidence for spin precession or tidal effects has been found.

Given that meaningful constraints on scalar-tensor gravity require observations of NSBH binaries, and the current number of confirmed NSBH events remains limited, the scope for testing such theories is still narrow. However, the prospects are promising: future gravitational wave observatories, both ground-based and space-based, including next generation detectors are expected to significantly increase the number of detected NSBH events. This will enable more stringent and statistically robust constraints on scalar-tensor modifications of gravity.
Furthermore, we anticipate the detection of systems with higher mass ratios, for which higher-order effects become increasingly prominent. These systems will enhance the sensitivity of gravitational wave measurements to deviations from general relativity, enabling more precise modeling of waveforms and stronger tests of alternative theories.

\acknowledgments
This work is supported by ``the Fundamental Research Funds for the Central Universities'' under the reference No. 2042024FG0009. The data, software, and/or web tools used in this research were obtained from the Gravitational Wave Open Science Center, a service of LIGO Laboratory, the LIGO Scientific Collaboration, and the Virgo Collaboration.

\appendix
\section{Robustness checks}
\label{robustness}
In some theory-agnostic tests, the prior for $\varphi_{-2}$ is chosen to include both negative and positive values. In our study, the prior for $\varphi_{-2}$ is restricted to negative values rather than extending to positive values. The rationale for this choice is that, for the specific scalar-tensor modified gravity theories we investigate, the mapping to $\varphi_{-2}$ fixes its sign. For example, when constraining the Brans-Dicke parameter $\omega_{\mathrm{BD}}$, we adopt the widest possible prior range, namely $\omega_{\mathrm{BD}} \in [0, +\infty)$. Correspondingly, we assume a uniform distribution for $1/(\omega_{\mathrm{BD}} + 2)$ in the interval $[0, 0.5]$, which maps to a uniform distribution for $\varphi_{-2}$ in $[-0.0027, 0]$. Since $\varphi_{-2}$ and $1/(\omega_{\mathrm{BD}} + 2)$ differ only by a constant factor, this can equivalently be regarded as performing parameter estimation for $1/(\omega_{\mathrm{BD}} + 2)$ within $[0, 0.5]$.
This prior choice is therefore motivated by the distinction between analyzing a specific modified-gravity theory and conducting a theory agnostic-analysis, where the latter typically adopts priors extending to both negative and positive values. To quantify the systematic error that would arise if one directly mapped the posterior from a theory-agnostic test with a symmetric prior (including both negative and positive values) into the scalar-tensor framework, we carried out parameter estimation tests comparing two cases:
	1.	Using a prior that spans both negative and positive values for $\varphi_{-2}
$, and then restricting the posterior to negative values afterward (the original posterior covering both negative and positive $\varphi_{-2}$ values is shown in Fig. 4(a), while the posterior truncated to negative values is shown in Fig. 4(b)).
2. Using a prior restricted to negative values from the outset (this is the approach adopted in our work).

\begin{figure}[H]
    \centering
    \begin{subfigure}[b]{0.45\textwidth}
        \centering
        \includegraphics[width=\textwidth]{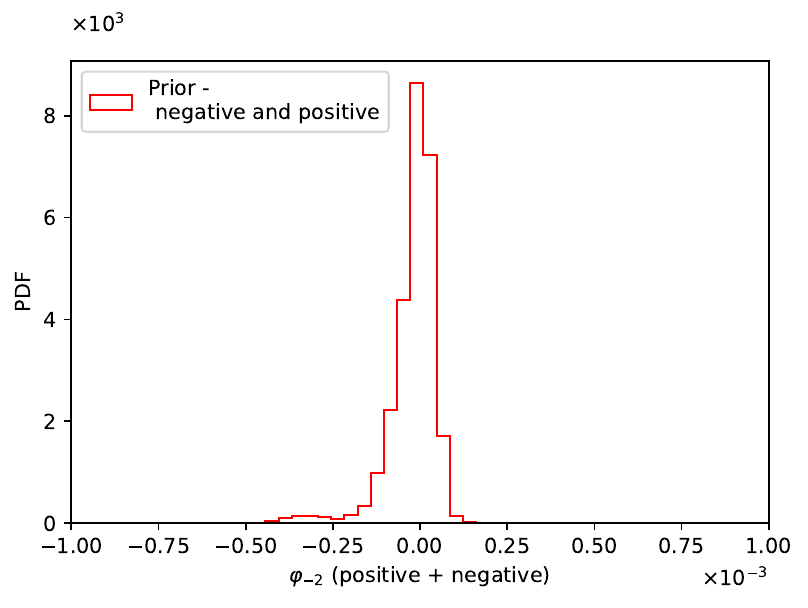}
        \caption{}
        \label{subfig:negative and positive}
    \end{subfigure}
    \hfill
    \begin{subfigure}[b]{0.45\textwidth}
        \centering
        \includegraphics[width=\textwidth]{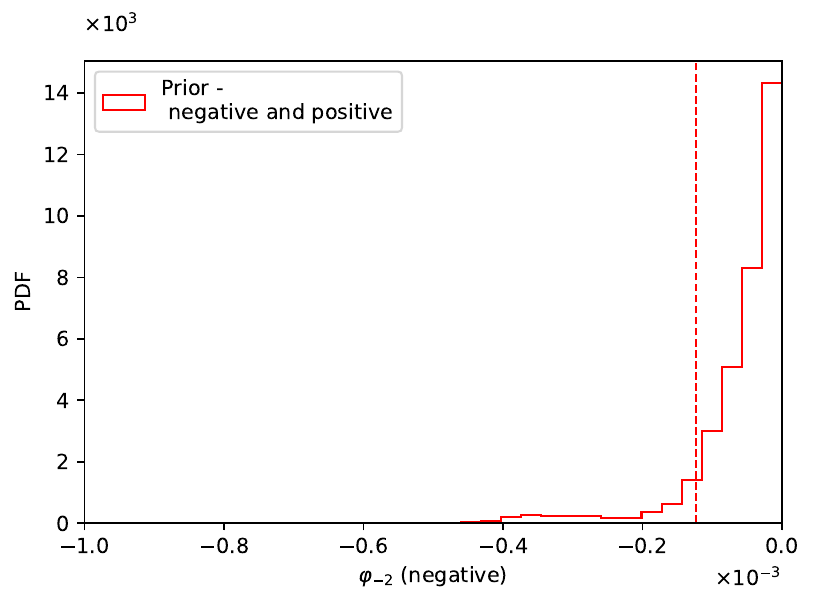}
        \caption{}
        \label{subfig:original}
    \end{subfigure}
    \caption{Prior including both negative and positive values}
    \label{fig:A1}
\end{figure}

We find that when the prior for $\varphi_{-2}$ spans both negative and positive values and the posterior is subsequently truncated to negative values, the lower bound of the 90\% credible interval is biased by about 25\% in our test, relative to using a negative-only prior for $\varphi_{-2}$. This underscores that, in scalar-tensor analyses, allowing both signs in the prior can introduce bias and should be avoided.

Our main analysis reconstructs the PSD directly from the strain using the Welch method, whereas some other works adopt the LVK event-specific PSDs. To assess robustness, we performed a test using the event-specific PSD provided in the LVK posterior sample releases. We include a discussion comparing the Bayesian parameter estimation results obtained with the two approaches: 1. the Welch method adopted in our work (Figure~5(a)) and 2. the event-specific PSDs from the LVK posterior sample releases (Figure~5(b)).

\begin{figure}[H]
    \centering
    \begin{subfigure}[b]{0.45\textwidth}
        \centering
        \includegraphics[width=\textwidth]{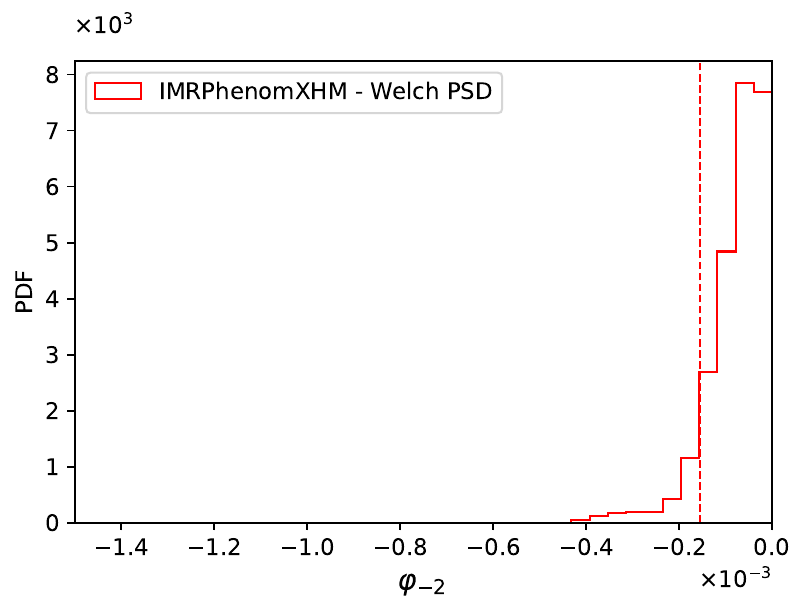}
        \caption{}
        \label{subfig:cut}
    \end{subfigure}
    \hfill
    \begin{subfigure}[b]{0.45\textwidth}
        \centering
        \includegraphics[width=\textwidth]{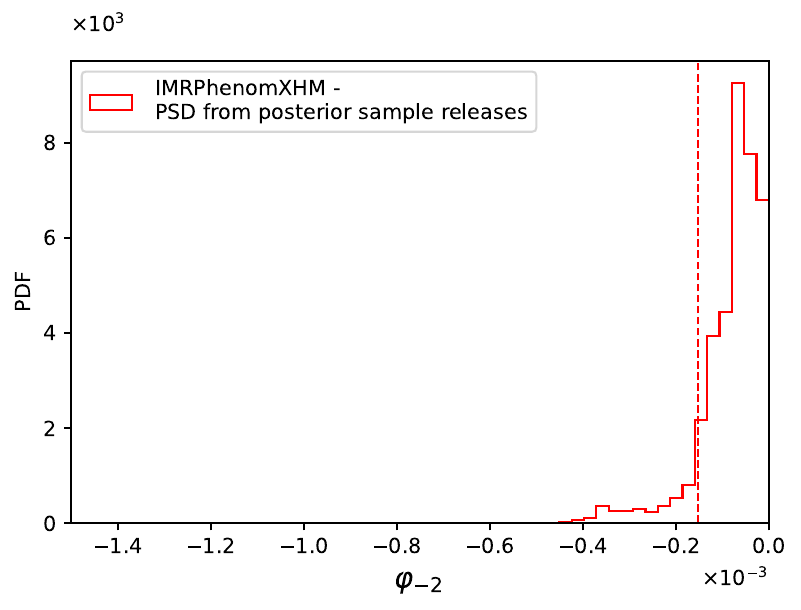}
        \caption{}
        \label{subfig:original}
    \end{subfigure}
    \caption{Comparison of results with different PSD choices}
    \label{fig:combined}
\end{figure}

We find that the posteriors of $\varphi_{-2}$ obtained from the two PSD choices show only minor differences, with the lower bound of the 90\% credible interval differing by no more than 5\%.

\bibliographystyle{JHEP}
\bibliography{Ref}

\providecommand{\href}[2]{#2}\begingroup\raggedright\begin{thebibliography}{10}

\bibitem{LIGOScientific:2014pky}
{\scshape LIGO Scientific} collaboration, \emph{{Advanced LIGO}}, \href{https://doi.org/10.1088/0264-9381/32/7/074001}{\emph{Class. Quant. Grav.} {\bfseries 32} (2015) 074001} [\href{https://arxiv.org/abs/1411.4547}{{\ttfamily 1411.4547}}].

\bibitem{VIRGO:2014yos}
{\scshape VIRGO} collaboration, \emph{{Advanced Virgo: a second-generation interferometric gravitational wave detector}}, \href{https://doi.org/10.1088/0264-9381/32/2/024001}{\emph{Class. Quant. Grav.} {\bfseries 32} (2015) 024001} [\href{https://arxiv.org/abs/1408.3978}{{\ttfamily 1408.3978}}].

\bibitem{LIGOScientific:2024elc}
{\scshape LIGO Scientific, Virgo,, KAGRA, VIRGO} collaboration, \emph{{Observation of Gravitational Waves from the Coalescence of a 2.5\textendash{}4.5 M$_{\odot}$ Compact Object and a Neutron Star}}, \href{https://doi.org/10.3847/2041-8213/ad5beb}{\emph{Astrophys. J. Lett.} {\bfseries 970} (2024) L34} [\href{https://arxiv.org/abs/2404.04248}{{\ttfamily 2404.04248}}].

\bibitem{LIGOScientific:2016lio}
{\scshape LIGO Scientific, Virgo} collaboration, \emph{{Tests of general relativity with GW150914}}, \href{https://doi.org/10.1103/PhysRevLett.116.221101}{\emph{Phys. Rev. Lett.} {\bfseries 116} (2016) 221101} [\href{https://arxiv.org/abs/1602.03841}{{\ttfamily 1602.03841}}].

\bibitem{LIGOScientific:2018dkp}
{\scshape LIGO Scientific, Virgo} collaboration, \emph{{Tests of General Relativity with GW170817}}, \href{https://doi.org/10.1103/PhysRevLett.123.011102}{\emph{Phys. Rev. Lett.} {\bfseries 123} (2019) 011102} [\href{https://arxiv.org/abs/1811.00364}{{\ttfamily 1811.00364}}].

\bibitem{Yunes:2013dva}
N.~Yunes and X.~Siemens, \emph{{Gravitational-Wave Tests of General Relativity with Ground-Based Detectors and Pulsar Timing-Arrays}}, \href{https://doi.org/10.12942/lrr-2013-9}{\emph{Living Rev. Rel.} {\bfseries 16} (2013) 9} [\href{https://arxiv.org/abs/1304.3473}{{\ttfamily 1304.3473}}].

\bibitem{Will:2014kxa}
C.~M. Will, \emph{{The Confrontation between General Relativity and Experiment}}, \href{https://doi.org/10.12942/lrr-2014-4}{\emph{Living Rev. Rel.} {\bfseries 17} (2014) 4} [\href{https://arxiv.org/abs/1403.7377}{{\ttfamily 1403.7377}}].

\bibitem{Berti:2015itd}
E.~Berti et~al., \emph{{Testing General Relativity with Present and Future Astrophysical Observations}}, \href{https://doi.org/10.1088/0264-9381/32/24/243001}{\emph{Class. Quant. Grav.} {\bfseries 32} (2015) 243001} [\href{https://arxiv.org/abs/1501.07274}{{\ttfamily 1501.07274}}].

\bibitem{Berti:2018cxi}
E.~Berti, K.~Yagi and N.~Yunes, \emph{{Extreme Gravity Tests with Gravitational Waves from Compact Binary Coalescences: (I) Inspiral-Merger}}, \href{https://doi.org/10.1007/s10714-018-2362-8}{\emph{Gen. Rel. Grav.} {\bfseries 50} (2018) 46} [\href{https://arxiv.org/abs/1801.03208}{{\ttfamily 1801.03208}}].

\bibitem{Sanger:2024axs}
E.~M. S{\"a}nger et~al., \emph{{Tests of General Relativity with GW230529: a neutron star merging with a lower mass-gap compact object}},  \href{https://arxiv.org/abs/2406.03568}{{\ttfamily 2406.03568}}.

\bibitem{Yunes:2016jcc}
N.~Yunes, K.~Yagi and F.~Pretorius, \emph{{Theoretical Physics Implications of the Binary Black-Hole Mergers GW150914 and GW151226}}, \href{https://doi.org/10.1103/PhysRevD.94.084002}{\emph{Phys. Rev. D} {\bfseries 94} (2016) 084002} [\href{https://arxiv.org/abs/1603.08955}{{\ttfamily 1603.08955}}].

\bibitem{Tahura:2019dgr}
S.~Tahura, K.~Yagi and Z.~Carson, \emph{{Testing Gravity with Gravitational Waves from Binary Black Hole Mergers: Contributions from Amplitude Corrections}}, \href{https://doi.org/10.1103/PhysRevD.100.104001}{\emph{Phys. Rev. D} {\bfseries 100} (2019) 104001} [\href{https://arxiv.org/abs/1907.10059}{{\ttfamily 1907.10059}}].

\bibitem{Nair:2019iur}
R.~Nair, S.~Perkins, H.~O. Silva and N.~Yunes, \emph{{Fundamental Physics Implications for Higher-Curvature Theories from Binary Black Hole Signals in the LIGO-Virgo Catalog GWTC-1}}, \href{https://doi.org/10.1103/PhysRevLett.123.191101}{\emph{Phys. Rev. Lett.} {\bfseries 123} (2019) 191101} [\href{https://arxiv.org/abs/1905.00870}{{\ttfamily 1905.00870}}].

\bibitem{Perkins:2021mhb}
S.~E. Perkins, R.~Nair, H.~O. Silva and N.~Yunes, \emph{{Improved gravitational-wave constraints on higher-order curvature theories of gravity}}, \href{https://doi.org/10.1103/PhysRevD.104.024060}{\emph{Phys. Rev. D} {\bfseries 104} (2021) 024060} [\href{https://arxiv.org/abs/2104.11189}{{\ttfamily 2104.11189}}].

\bibitem{Wang:2023wgv}
B.~Wang, C.~Shi, J.-d. Zhang, Y.-M. hu and J.~Mei, \emph{{Constraining the Einstein-dilaton-Gauss-Bonnet theory with higher harmonics and the merger-ringdown contribution using GWTC-3}}, \href{https://doi.org/10.1103/PhysRevD.108.044061}{\emph{Phys. Rev. D} {\bfseries 108} (2023) 044061} [\href{https://arxiv.org/abs/2302.10112}{{\ttfamily 2302.10112}}].

\bibitem{Gao:2024rel}
B.~Gao, S.-P. Tang, H.-T. Wang, J.~Yan and Y.-Z. Fan, \emph{{Constraints on Einstein-dilation-Gauss-Bonnet gravity and the electric charge of compact binary systems from GW230529}}, \href{https://doi.org/10.1103/PhysRevD.110.044022}{\emph{Phys. Rev. D} {\bfseries 110} (2024) 044022} [\href{https://arxiv.org/abs/2405.13279}{{\ttfamily 2405.13279}}].

\bibitem{Vijaykumar:2020nzc}
A.~Vijaykumar, S.~J. Kapadia and P.~Ajith, \emph{{Constraints on the time variation of the gravitational constant using gravitational-wave observations of binary neutron stars}}, \href{https://doi.org/10.1103/PhysRevLett.126.141104}{\emph{Phys. Rev. Lett.} {\bfseries 126} (2021) 141104} [\href{https://arxiv.org/abs/2003.12832}{{\ttfamily 2003.12832}}].

\bibitem{Zhao:2019suc}
J.~Zhao, L.~Shao, Z.~Cao and B.-Q. Ma, \emph{{Reduced-order surrogate models for scalar-tensor gravity in the strong field regime and applications to binary pulsars and GW170817}}, \href{https://doi.org/10.1103/PhysRevD.100.064034}{\emph{Phys. Rev. D} {\bfseries 100} (2019) 064034} [\href{https://arxiv.org/abs/1907.00780}{{\ttfamily 1907.00780}}].

\bibitem{Niu:2021nic}
R.~Niu, X.~Zhang, B.~Wang and W.~Zhao, \emph{{Constraining Scalar-tensor Theories Using Neutron Star\textendash{}Black Hole Gravitational Wave Events}}, \href{https://doi.org/10.3847/1538-4357/ac1d4f}{\emph{Astrophys. J.} {\bfseries 921} (2021) 149} [\href{https://arxiv.org/abs/2105.13644}{{\ttfamily 2105.13644}}].

\bibitem{Damour:1992we}
T.~Damour and G.~Esposito-Farese, \emph{{Tensor multiscalar theories of gravitation}}, \href{https://doi.org/10.1088/0264-9381/9/9/015}{\emph{Class. Quant. Grav.} {\bfseries 9} (1992) 2093}.

\bibitem{sotiriou2014gravity}
T.~P. Sotiriou, \emph{Gravity and scalar fields},  in \emph{Modifications of Einstein's Theory of Gravity at Large Distances}, pp.~3--24, Springer, (2014).

\bibitem{Bernard:2022noq}
L.~Bernard, L.~Blanchet and D.~Trestini, \emph{{Gravitational waves in scalar-tensor theory to one-and-a-half post-Newtonian order}}, \href{https://doi.org/10.1088/1475-7516/2022/08/008}{\emph{JCAP} {\bfseries 08} (2022) 008} [\href{https://arxiv.org/abs/2201.10924}{{\ttfamily 2201.10924}}].

\bibitem{Sennett:2016klh}
N.~Sennett, S.~Marsat and A.~Buonanno, \emph{{Gravitational waveforms in scalar-tensor gravity at 2PN relative order}}, \href{https://doi.org/10.1103/PhysRevD.94.084003}{\emph{Phys. Rev. D} {\bfseries 94} (2016) 084003} [\href{https://arxiv.org/abs/1607.01420}{{\ttfamily 1607.01420}}].

\bibitem{Trestini:2024zpi}
D.~Trestini, \emph{{Quasi-Keplerian parametrization for eccentric compact binaries in scalar-tensor theories at second post-Newtonian order and applications}}, \href{https://doi.org/10.1103/PhysRevD.109.104003}{\emph{Phys. Rev. D} {\bfseries 109} (2024) 104003} [\href{https://arxiv.org/abs/2401.06844}{{\ttfamily 2401.06844}}].

\bibitem{Trestini:2024mfs}
D.~Trestini, \emph{{Gravitational waves from quasielliptic compact binaries in scalar-tensor theory to one-and-a-half post-Newtonian order}},  \href{https://arxiv.org/abs/2410.12898}{{\ttfamily 2410.12898}}.

\bibitem{Bhattacharyya:2024aeq}
A.~Bhattacharyya, D.~Ghosh, S.~Ghosh and S.~Pal, \emph{{Observables from classical black hole scattering in Scalar-Tensor theory of gravity from worldline quantum field theory}}, \href{https://doi.org/10.1007/JHEP04(2024)015}{\emph{JHEP} {\bfseries 04} (2024) 015} [\href{https://arxiv.org/abs/2401.05492}{{\ttfamily 2401.05492}}].

\bibitem{polchinski1998string}
J.~G. Polchinski, \emph{String theory, volume I: An introduction to the bosonic string}. Cambridge university press Cambridge, 1998.

\bibitem{duff1994kaluza}
M.~J. Duff, \emph{Kaluza-klein theory in perspective},  in \emph{Proc. of the Symposium: The Oskar Klein Centenary, World Scientific, Singapore}, pp.~22--35, World Scientific, 1994.

\bibitem{Randall:1999ee}
L.~Randall and R.~Sundrum, \emph{{A Large mass hierarchy from a small extra dimension}}, \href{https://doi.org/10.1103/PhysRevLett.83.3370}{\emph{Phys. Rev. Lett.} {\bfseries 83} (1999) 3370} [\href{https://arxiv.org/abs/hep-ph/9905221}{{\ttfamily hep-ph/9905221}}].

\bibitem{Randall:1999vf}
L.~Randall and R.~Sundrum, \emph{{An Alternative to compactification}}, \href{https://doi.org/10.1103/PhysRevLett.83.4690}{\emph{Phys. Rev. Lett.} {\bfseries 83} (1999) 4690} [\href{https://arxiv.org/abs/hep-th/9906064}{{\ttfamily hep-th/9906064}}].

\bibitem{Riazuelo:2001mg}
A.~Riazuelo and J.-P. Uzan, \emph{{Cosmological observations in scalar - tensor quintessence}}, \href{https://doi.org/10.1103/PhysRevD.66.023525}{\emph{Phys. Rev. D} {\bfseries 66} (2002) 023525} [\href{https://arxiv.org/abs/astro-ph/0107386}{{\ttfamily astro-ph/0107386}}].

\bibitem{Brax:2004qh}
P.~Brax, C.~van~de Bruck, A.-C. Davis, J.~Khoury and A.~Weltman, \emph{{Detecting dark energy in orbit: The cosmological chameleon}}, \href{https://doi.org/10.1103/PhysRevD.70.123518}{\emph{Phys. Rev. D} {\bfseries 70} (2004) 123518} [\href{https://arxiv.org/abs/astro-ph/0408415}{{\ttfamily astro-ph/0408415}}].

\bibitem{Kainulainen:2004vk}
K.~Kainulainen and D.~Sunhede, \emph{{Dark energy from large extra dimensions}}, \href{https://doi.org/10.1103/PhysRevD.73.083510}{\emph{Phys. Rev. D} {\bfseries 73} (2006) 083510} [\href{https://arxiv.org/abs/astro-ph/0412609}{{\ttfamily astro-ph/0412609}}].

\bibitem{Clifton:2011jh}
T.~Clifton, P.~G. Ferreira, A.~Padilla and C.~Skordis, \emph{{Modified Gravity and Cosmology}}, \href{https://doi.org/10.1016/j.physrep.2012.01.001}{\emph{Phys. Rept.} {\bfseries 513} (2012) 1} [\href{https://arxiv.org/abs/1106.2476}{{\ttfamily 1106.2476}}].

\bibitem{Burd:1991ns}
A.~Burd and A.~Coley, \emph{{Extended inflation and generalized scalar - tensor theories}}, \href{https://doi.org/10.1016/0370-2693(91)90941-I}{\emph{Phys. Lett. B} {\bfseries 267} (1991) 330}.

\bibitem{Barrow:1990nv}
J.~D. Barrow and K.-i. Maeda, \emph{{Extended inflationary universes}}, \href{https://doi.org/10.1016/0550-3213(90)90272-F}{\emph{Nucl. Phys. B} {\bfseries 341} (1990) 294}.

\bibitem{Brax:2005ew}
P.~Brax, C.~van~de Bruck, A.-C. Davis and A.~M. Green, \emph{{Small scale structure formation in chameleon cosmology}}, \href{https://doi.org/10.1016/j.physletb.2005.12.055}{\emph{Phys. Lett. B} {\bfseries 633} (2006) 441} [\href{https://arxiv.org/abs/astro-ph/0509878}{{\ttfamily astro-ph/0509878}}].

\bibitem{Coc:2006rt}
A.~Coc, K.~A. Olive, J.-P. Uzan and E.~Vangioni, \emph{{Big bang nucleosynthesis constraints on scalar-tensor theories of gravity}}, \href{https://doi.org/10.1103/PhysRevD.73.083525}{\emph{Phys. Rev. D} {\bfseries 73} (2006) 083525} [\href{https://arxiv.org/abs/astro-ph/0601299}{{\ttfamily astro-ph/0601299}}].

\bibitem{Damour:1998ae}
T.~Damour and B.~Pichon, \emph{{Big bang nucleosynthesis and tensor - scalar gravity}}, \href{https://doi.org/10.1103/PhysRevD.59.123502}{\emph{Phys. Rev. D} {\bfseries 59} (1999) 123502} [\href{https://arxiv.org/abs/astro-ph/9807176}{{\ttfamily astro-ph/9807176}}].

\bibitem{Larena:2005tu}
J.~Larena, J.-M. Alimi and A.~Serna, \emph{{Big Bang nucleosynthesis in scalar tensor gravity: The key problem of the primordial Li-7 abundance}}, \href{https://doi.org/10.1086/511028}{\emph{Astrophys. J.} {\bfseries 658} (2007) 1} [\href{https://arxiv.org/abs/astro-ph/0511693}{{\ttfamily astro-ph/0511693}}].

\bibitem{Torres:1995je}
D.~F. Torres, \emph{{Nucleosynthesis bounds on scalar tensor gravity: power law couplings}}, \href{https://doi.org/10.1016/0370-2693(95)01098-B}{\emph{Phys. Lett. B} {\bfseries 359} (1995) 249}.

\bibitem{Khoury:2010xi}
J.~Khoury, \emph{{Theories of Dark Energy with Screening Mechanisms}},  \href{https://arxiv.org/abs/1011.5909}{{\ttfamily 1011.5909}}.

\bibitem{Khoury:2003rn}
J.~Khoury and A.~Weltman, \emph{{Chameleon cosmology}}, \href{https://doi.org/10.1103/PhysRevD.69.044026}{\emph{Phys. Rev. D} {\bfseries 69} (2004) 044026} [\href{https://arxiv.org/abs/astro-ph/0309411}{{\ttfamily astro-ph/0309411}}].

\bibitem{Vainshtein:1972sx}
A.~I. Vainshtein, \emph{{To the problem of nonvanishing gravitation mass}}, \href{https://doi.org/10.1016/0370-2693(72)90147-5}{\emph{Phys. Lett. B} {\bfseries 39} (1972) 393}.

\bibitem{Babichev:2013usa}
E.~Babichev and C.~Deffayet, \emph{{An introduction to the Vainshtein mechanism}}, \href{https://doi.org/10.1088/0264-9381/30/18/184001}{\emph{Class. Quant. Grav.} {\bfseries 30} (2013) 184001} [\href{https://arxiv.org/abs/1304.7240}{{\ttfamily 1304.7240}}].

\bibitem{Hinterbichler:2010es}
K.~Hinterbichler and J.~Khoury, \emph{{Symmetron Fields: Screening Long-Range Forces Through Local Symmetry Restoration}}, \href{https://doi.org/10.1103/PhysRevLett.104.231301}{\emph{Phys. Rev. Lett.} {\bfseries 104} (2010) 231301} [\href{https://arxiv.org/abs/1001.4525}{{\ttfamily 1001.4525}}].

\bibitem{Joyce:2014kja}
A.~Joyce, B.~Jain, J.~Khoury and M.~Trodden, \emph{{Beyond the Cosmological Standard Model}}, \href{https://doi.org/10.1016/j.physrep.2014.12.002}{\emph{Phys. Rept.} {\bfseries 568} (2015) 1} [\href{https://arxiv.org/abs/1407.0059}{{\ttfamily 1407.0059}}].

\bibitem{Brax:2012bsa}
P.~Brax, \emph{{Screened modified gravity}}, \href{https://doi.org/10.5506/APhysPolB.43.2307}{\emph{Acta Phys. Polon. B} {\bfseries 43} (2012) 2307} [\href{https://arxiv.org/abs/1211.5237}{{\ttfamily 1211.5237}}].

\bibitem{Fischer:2024eic}
H.~Fischer, C.~K\"ading and M.~Pitschmann, \emph{{Screened Scalar Fields in the Laboratory and the Solar System}}, \href{https://doi.org/10.3390/universe10070297}{\emph{Universe} {\bfseries 10} (2024) 297} [\href{https://arxiv.org/abs/2405.14638}{{\ttfamily 2405.14638}}].

\bibitem{Brans:1961sx}
C.~Brans and R.~H. Dicke, \emph{{Mach's principle and a relativistic theory of gravitation}}, \href{https://doi.org/10.1103/PhysRev.124.925}{\emph{Phys. Rev.} {\bfseries 124} (1961) 925}.

\bibitem{Freire:2012mg}
P.~C.~C. Freire, N.~Wex, G.~Esposito-Farese, J.~P.~W. Verbiest, M.~Bailes, B.~A. Jacoby et~al., \emph{{The relativistic pulsar-white dwarf binary PSR J1738+0333 II. The most stringent test of scalar-tensor gravity}}, \href{https://doi.org/10.1111/j.1365-2966.2012.21253.x}{\emph{Mon. Not. Roy. Astron. Soc.} {\bfseries 423} (2012) 3328} [\href{https://arxiv.org/abs/1205.1450}{{\ttfamily 1205.1450}}].

\bibitem{Bertotti:2003rm}
B.~Bertotti, L.~Iess and P.~Tortora, \emph{{A test of general relativity using radio links with the Cassini spacecraft}}, \href{https://doi.org/10.1038/nature01997}{\emph{Nature} {\bfseries 425} (2003) 374}.

\bibitem{Berti:2004bd}
E.~Berti, A.~Buonanno and C.~M. Will, \emph{{Estimating spinning binary parameters and testing alternative theories of gravity with LISA}}, \href{https://doi.org/10.1103/PhysRevD.71.084025}{\emph{Phys. Rev. D} {\bfseries 71} (2005) 084025} [\href{https://arxiv.org/abs/gr-qc/0411129}{{\ttfamily gr-qc/0411129}}].

\bibitem{Yagi:2009zm}
K.~Yagi and T.~Tanaka, \emph{{Constraining alternative theories of gravity by gravitational waves from precessing eccentric compact binaries with LISA}}, \href{https://doi.org/10.1103/PhysRevD.81.109902}{\emph{Phys. Rev. D} {\bfseries 81} (2010) 064008} [\href{https://arxiv.org/abs/0906.4269}{{\ttfamily 0906.4269}}].

\bibitem{Scharre:2001hn}
P.~D. Scharre and C.~M. Will, \emph{{Testing scalar tensor gravity using space gravitational wave interferometers}}, \href{https://doi.org/10.1103/PhysRevD.65.042002}{\emph{Phys. Rev. D} {\bfseries 65} (2002) 042002} [\href{https://arxiv.org/abs/gr-qc/0109044}{{\ttfamily gr-qc/0109044}}].

\bibitem{Jiang:2021htl}
T.~Jiang, N.~Dai, Y.~Gong, D.~Liang and C.~Zhang, \emph{{Constraint on Brans-Dicke theory from intermediate/extreme mass ratio inspirals}}, \href{https://doi.org/10.1088/1475-7516/2022/12/023}{\emph{JCAP} {\bfseries 12} (2022) 023} [\href{https://arxiv.org/abs/2107.02700}{{\ttfamily 2107.02700}}].

\bibitem{Gao:2022hsn}
Q.~Gao, Y.~You, Y.~Gong, C.~Zhang and C.~Zhang, \emph{{Testing alternative theories of gravity with space-based gravitational wave detectors}}, \href{https://doi.org/10.1103/PhysRevD.108.024027}{\emph{Phys. Rev. D} {\bfseries 108} (2023) 024027} [\href{https://arxiv.org/abs/2212.03789}{{\ttfamily 2212.03789}}].

\bibitem{Ashton:2018jfp}
G.~Ashton et~al., \emph{{BILBY: A user-friendly Bayesian inference library for gravitational-wave astronomy}}, \href{https://doi.org/10.3847/1538-4365/ab06fc}{\emph{Astrophys. J. Suppl.} {\bfseries 241} (2019) 27} [\href{https://arxiv.org/abs/1811.02042}{{\ttfamily 1811.02042}}].

\bibitem{Tan:2023fyl}
J.~Tan and B.~Wang, \emph{{Constraints on Brans-Dicke gravity from neutron star-black hole merger events using higher harmonics}}, \href{https://doi.org/10.1103/PhysRevD.109.084036}{\emph{Phys. Rev. D} {\bfseries 109} (2024) 084036} [\href{https://arxiv.org/abs/2312.07017}{{\ttfamily 2312.07017}}].

\bibitem{Takeda:2023wqn}
H.~Takeda, S.~Tsujikawa and A.~Nishizawa, \emph{{Gravitational-wave constraints on scalar-tensor gravity from a neutron star and black-hole binary GW200115}},  \href{https://arxiv.org/abs/2311.09281}{{\ttfamily 2311.09281}}.

\bibitem{Biwer:2018osg}
C.~M. Biwer, C.~D. Capano, S.~De, M.~Cabero, D.~A. Brown, A.~H. Nitz et~al., \emph{{PyCBC Inference: A Python-based parameter estimation toolkit for compact binary coalescence signals}}, \href{https://doi.org/10.1088/1538-3873/aaef0b}{\emph{Publ. Astron. Soc. Pac.} {\bfseries 131} (2019) 024503} [\href{https://arxiv.org/abs/1807.10312}{{\ttfamily 1807.10312}}].

\bibitem{Thrane:2018qnx}
E.~Thrane and C.~Talbot, \emph{{An introduction to Bayesian inference in gravitational-wave astronomy: parameter estimation, model selection, and hierarchical models}}, \href{https://doi.org/10.1017/pasa.2019.2}{\emph{Publ. Astron. Soc. Austral.} {\bfseries 36} (2019) e010} [\href{https://arxiv.org/abs/1809.02293}{{\ttfamily 1809.02293}}].

\bibitem{Smith:2021bqc}
R.~Smith et~al., \emph{{Bayesian Inference for Gravitational Waves from Binary Neutron Star Mergers in Third Generation Observatories}}, \href{https://doi.org/10.1103/PhysRevLett.127.081102}{\emph{Phys. Rev. Lett.} {\bfseries 127} (2021) 081102} [\href{https://arxiv.org/abs/2103.12274}{{\ttfamily 2103.12274}}].

\bibitem{Lyu:2022gdr}
Z.~Lyu, N.~Jiang and K.~Yagi, \emph{{Constraints on Einstein-dilation-Gauss-Bonnet gravity from black hole-neutron star gravitational wave events}}, \href{https://doi.org/10.1103/PhysRevD.105.064001}{\emph{Phys. Rev. D} {\bfseries 105} (2022) 064001} [\href{https://arxiv.org/abs/2201.02543}{{\ttfamily 2201.02543}}].

\bibitem{Yunes:2009ke}
N.~Yunes and F.~Pretorius, \emph{{Fundamental Theoretical Bias in Gravitational Wave Astrophysics and the Parameterized Post-Einsteinian Framework}}, \href{https://doi.org/10.1103/PhysRevD.80.122003}{\emph{Phys. Rev. D} {\bfseries 80} (2009) 122003} [\href{https://arxiv.org/abs/0909.3328}{{\ttfamily 0909.3328}}].

\bibitem{Yunes:2010qb}
N.~Yunes and S.~A. Hughes, \emph{{Binary Pulsar Constraints on the Parameterized post-Einsteinian Framework}}, \href{https://doi.org/10.1103/PhysRevD.82.082002}{\emph{Phys. Rev. D} {\bfseries 82} (2010) 082002} [\href{https://arxiv.org/abs/1007.1995}{{\ttfamily 1007.1995}}].

\bibitem{Mirshekari:2011yq}
S.~Mirshekari, N.~Yunes and C.~M. Will, \emph{{Constraining Generic Lorentz Violation and the Speed of the Graviton with Gravitational Waves}}, \href{https://doi.org/10.1103/PhysRevD.85.024041}{\emph{Phys. Rev. D} {\bfseries 85} (2012) 024041} [\href{https://arxiv.org/abs/1110.2720}{{\ttfamily 1110.2720}}].

\bibitem{Tahura:2018zuq}
S.~Tahura and K.~Yagi, \emph{{Parameterized Post-Einsteinian Gravitational Waveforms in Various Modified Theories of Gravity}}, \href{https://doi.org/10.1103/PhysRevD.98.084042}{\emph{Phys. Rev. D} {\bfseries 98} (2018) 084042} [\href{https://arxiv.org/abs/1809.00259}{{\ttfamily 1809.00259}}].

\bibitem{Blanchet:2013haa}
L.~Blanchet, \emph{{Gravitational Radiation from Post-Newtonian Sources and Inspiralling Compact Binaries}}, \href{https://doi.org/10.12942/lrr-2014-2}{\emph{Living Rev. Rel.} {\bfseries 17} (2014) 2} [\href{https://arxiv.org/abs/1310.1528}{{\ttfamily 1310.1528}}].

\bibitem{Yunes:2009bv}
N.~Yunes, F.~Pretorius and D.~Spergel, \emph{{Constraining the evolutionary history of Newton's constant with gravitational wave observations}}, \href{https://doi.org/10.1103/PhysRevD.81.064018}{\emph{Phys. Rev. D} {\bfseries 81} (2010) 064018} [\href{https://arxiv.org/abs/0912.2724}{{\ttfamily 0912.2724}}].

\bibitem{Mezzasoma:2022pjb}
S.~Mezzasoma and N.~Yunes, \emph{{Theory-agnostic framework for inspiral tests of general relativity with higher-harmonic gravitational waves}}, \href{https://doi.org/10.1103/PhysRevD.106.024026}{\emph{Phys. Rev. D} {\bfseries 106} (2022) 024026} [\href{https://arxiv.org/abs/2203.15934}{{\ttfamily 2203.15934}}].

\bibitem{Chiba:1997ms}
T.~Chiba, T.~Harada and K.-i. Nakao, \emph{{Gravitational physics in scalar tensor theories: Tests of strong field gravity}}, \href{https://doi.org/10.1143/PTPS.128.335}{\emph{Prog. Theor. Phys. Suppl.} {\bfseries 128} (1997) 335}.

\bibitem{Will:1994fb}
C.~M. Will, \emph{{Testing scalar - tensor gravity with gravitational wave observations of inspiraling compact binaries}}, \href{https://doi.org/10.1103/PhysRevD.50.6058}{\emph{Phys. Rev. D} {\bfseries 50} (1994) 6058} [\href{https://arxiv.org/abs/gr-qc/9406022}{{\ttfamily gr-qc/9406022}}].

\bibitem{Zhang:2017sym}
X.~Zhang, J.~Yu, T.~Liu, W.~Zhao and A.~Wang, \emph{{Testing Brans-Dicke gravity using the Einstein telescope}}, \href{https://doi.org/10.1103/PhysRevD.95.124008}{\emph{Phys. Rev. D} {\bfseries 95} (2017) 124008} [\href{https://arxiv.org/abs/1703.09853}{{\ttfamily 1703.09853}}].

\bibitem{Liu:2020moh}
T.~Liu, W.~Zhao and Y.~Wang, \emph{{Gravitational waveforms from the quasicircular inspiral of compact binaries in massive Brans-Dicke theory}}, \href{https://doi.org/10.1103/PhysRevD.102.124035}{\emph{Phys. Rev. D} {\bfseries 102} (2020) 124035} [\href{https://arxiv.org/abs/2007.10068}{{\ttfamily 2007.10068}}].

\bibitem{Hawking:1972qk}
S.~W. Hawking, \emph{{Black holes in the Brans-Dicke theory of gravitation}}, \href{https://doi.org/10.1007/BF01877518}{\emph{Commun. Math. Phys.} {\bfseries 25} (1972) 167}.

\bibitem{Bekenstein:1995un}
J.~D. Bekenstein, \emph{{Novel \textquoteleft{}\textquoteleft{}no-scalar-hair\textquoteright{}\textquoteright{} theorem for black holes}}, \href{https://doi.org/10.1103/PhysRevD.51.R6608}{\emph{Phys. Rev. D} {\bfseries 51} (1995) R6608}.

\bibitem{Sotiriou:2011dz}
T.~P. Sotiriou and V.~Faraoni, \emph{{Black holes in scalar-tensor gravity}}, \href{https://doi.org/10.1103/PhysRevLett.108.081103}{\emph{Phys. Rev. Lett.} {\bfseries 108} (2012) 081103} [\href{https://arxiv.org/abs/1109.6324}{{\ttfamily 1109.6324}}].

\bibitem{Sampson:2013jpa}
L.~Sampson, N.~Cornish and N.~Yunes, \emph{{Mismodeling in gravitational-wave astronomy: The trouble with templates}}, \href{https://doi.org/10.1103/PhysRevD.89.064037}{\emph{Phys. Rev. D} {\bfseries 89} (2014) 064037} [\href{https://arxiv.org/abs/1311.4898}{{\ttfamily 1311.4898}}].

\bibitem{LIGOScientific:2019fpa}
{\scshape LIGO Scientific, Virgo} collaboration, \emph{{Tests of General Relativity with the Binary Black Hole Signals from the LIGO-Virgo Catalog GWTC-1}}, \href{https://doi.org/10.1103/PhysRevD.100.104036}{\emph{Phys. Rev. D} {\bfseries 100} (2019) 104036} [\href{https://arxiv.org/abs/1903.04467}{{\ttfamily 1903.04467}}].

\bibitem{Khan:2015jqa}
S.~Khan, S.~Husa, M.~Hannam, F.~Ohme, M.~P{\"u}rrer, X.~Jim{\'e}nez~Forteza et~al., \emph{{Frequency-domain gravitational waves from nonprecessing black-hole binaries. II. A phenomenological model for the advanced detector era}}, \href{https://doi.org/10.1103/PhysRevD.93.044007}{\emph{Phys. Rev. D} {\bfseries 93} (2016) 044007} [\href{https://arxiv.org/abs/1508.07253}{{\ttfamily 1508.07253}}].

\bibitem{Foreman-Mackey:2012any}
D.~Foreman-Mackey, D.~W. Hogg, D.~Lang and J.~Goodman, \emph{{emcee: The MCMC Hammer}}, \href{https://doi.org/10.1086/670067}{\emph{Publ. Astron. Soc. Pac.} {\bfseries 125} (2013) 306} [\href{https://arxiv.org/abs/1202.3665}{{\ttfamily 1202.3665}}].

\bibitem{LIGOScientific:2021qlt}
{\scshape LIGO Scientific, KAGRA, VIRGO} collaboration, \emph{{Observation of Gravitational Waves from Two Neutron Star\textendash{}Black Hole Coalescences}}, \href{https://doi.org/10.3847/2041-8213/ac082e}{\emph{Astrophys. J. Lett.} {\bfseries 915} (2021) L5} [\href{https://arxiv.org/abs/2106.15163}{{\ttfamily 2106.15163}}].

\bibitem{collaboration2018ligo}
L.~S. Collaboration, \emph{Ligo algorithm library—lalsuite, free software (gpl), doi: 10.7935}, \href{https://doi.org/10.7935/GT1W-FZ16}{\emph{GT1W-FZ16} (2018) }.

\bibitem{deRham:2012fw}
C.~de~Rham, A.~J. Tolley and D.~H. Wesley, \emph{{Vainshtein Mechanism in Binary Pulsars}}, \href{https://doi.org/10.1103/PhysRevD.87.044025}{\emph{Phys. Rev. D} {\bfseries 87} (2013) 044025} [\href{https://arxiv.org/abs/1208.0580}{{\ttfamily 1208.0580}}].

\bibitem{Brax:2012jr}
P.~Brax, C.~Burrage and A.-C. Davis, \emph{{Screening fifth forces in k-essence and DBI models}}, \href{https://doi.org/10.1088/1475-7516/2013/01/020}{\emph{JCAP} {\bfseries 01} (2013) 020} [\href{https://arxiv.org/abs/1209.1293}{{\ttfamily 1209.1293}}].

\bibitem{Babichev:2009ee}
E.~Babichev, C.~Deffayet and R.~Ziour, \emph{{k-Mouflage gravity}}, \href{https://doi.org/10.1142/S0218271809016107}{\emph{Int. J. Mod. Phys. D} {\bfseries 18} (2009) 2147} [\href{https://arxiv.org/abs/0905.2943}{{\ttfamily 0905.2943}}].

\bibitem{Isi:2019asy}
M.~Isi, K.~Chatziioannou and W.~M. Farr, \emph{{Hierarchical test of general relativity with gravitational waves}}, \href{https://doi.org/10.1103/PhysRevLett.123.121101}{\emph{Phys. Rev. Lett.} {\bfseries 123} (2019) 121101} [\href{https://arxiv.org/abs/1904.08011}{{\ttfamily 1904.08011}}].

\bibitem{Isi:2022cii}
M.~Isi, W.~M. Farr and K.~Chatziioannou, \emph{{Comparing Bayes factors and hierarchical inference for testing general relativity with gravitational waves}}, \href{https://doi.org/10.1103/PhysRevD.106.024048}{\emph{Phys. Rev. D} {\bfseries 106} (2022) 024048} [\href{https://arxiv.org/abs/2204.10742}{{\ttfamily 2204.10742}}].

\bibitem{Chandramouli:2024vhw}
R.~S. Chandramouli, K.~Prokup, E.~Berti and N.~Yunes, \emph{{Systematic biases due to waveform mismodeling in parametrized post-Einsteinian tests of general relativity: The impact of neglecting spin precession and higher modes}}, \href{https://doi.org/10.1103/PhysRevD.111.044026}{\emph{Phys. Rev. D} {\bfseries 111} (2025) 044026} [\href{https://arxiv.org/abs/2410.06254}{{\ttfamily 2410.06254}}].

\bibitem{Owen:2023mid}
C.~B. Owen, C.-J. Haster, S.~Perkins, N.~J. Cornish and N.~Yunes, \emph{{Waveform accuracy and systematic uncertainties in current gravitational wave observations}}, \href{https://doi.org/10.1103/PhysRevD.108.044018}{\emph{Phys. Rev. D} {\bfseries 108} (2023) 044018} [\href{https://arxiv.org/abs/2301.11941}{{\ttfamily 2301.11941}}].

\bibitem{Read:2023hkv}
J.~S. Read, \emph{{Waveform uncertainty quantification and interpretation for gravitational-wave astronomy}}, \href{https://doi.org/10.1088/1361-6382/acd29b}{\emph{Class. Quant. Grav.} {\bfseries 40} (2023) 135002} [\href{https://arxiv.org/abs/2301.06630}{{\ttfamily 2301.06630}}].

\bibitem{Wade:2014vqa}
L.~Wade, J.~D.~E. Creighton, E.~Ochsner, B.~D. Lackey, B.~F. Farr, T.~B. Littenberg et~al., \emph{{Systematic and statistical errors in a bayesian approach to the estimation of the neutron-star equation of state using advanced gravitational wave detectors}}, \href{https://doi.org/10.1103/PhysRevD.89.103012}{\emph{Phys. Rev. D} {\bfseries 89} (2014) 103012} [\href{https://arxiv.org/abs/1402.5156}{{\ttfamily 1402.5156}}].

\bibitem{Favata:2013rwa}
M.~Favata, \emph{{Systematic parameter errors in inspiraling neutron star binaries}}, \href{https://doi.org/10.1103/PhysRevLett.112.101101}{\emph{Phys. Rev. Lett.} {\bfseries 112} (2014) 101101} [\href{https://arxiv.org/abs/1310.8288}{{\ttfamily 1310.8288}}].

\end{thebibliography}\endgroup

\end{document}